\documentclass[aps,prd,preprintnumbers,amsmath,amssymb,nofootinbib,11pt]{revtex4}
\usepackage{eepic}
\usepackage{indentfirst}
\usepackage{mathrsfs}
\usepackage{fancyhdr}
\usepackage{ulem}
\usepackage{float}
\usepackage{graphicx}
\usepackage[
colorlinks=true, linkcolor=black, breaklinks=true, urlcolor=blue,
citecolor=green]{hyperref}
\usepackage{epstopdf}
\usepackage{bm,bbm}

\usepackage{tabularx}
\usepackage{subfigure}
\usepackage{epsfig}
\usepackage{color}
\usepackage{slashed}
\usepackage{hyperref}

\newcommand{\be}{\begin{equation}}
\newcommand{\ee}{\end{equation}}
\newcommand{\bqa}{\begin{eqnarray}}
\newcommand{\eqa}{\end{eqnarray}}
\renewcommand{\L}{\mathscr{L}}

\newcommand{\bra}{\langle}
\newcommand{\ket}{\rangle}
\newcommand{\nn}{\nonumber}
\newcommand{\MeV}{\,\text{MeV}}
\newcommand{\GeV}{\,\text{GeV}}
\renewcommand{\vec}[1]{\mathbf{#1}}

\newcommand{\qperp}{\vec q_\perp}
\newcommand{\q}{\vec q}

\begin{document}
\pagestyle{plain}

\title {\boldmath  Predictions of $\Upsilon(4S) \to h_b(1P,2P) \pi^+\pi^-$ transitions}

\author{ Yun-Hua~Chen}\email{yhchen@ustb.edu.cn}
\affiliation{School of Mathematics and Physics, University
of Science and Technology Beijing, Beijing 100083, China}

\begin{abstract}

In this work, we study the contributions of the intermediate bottomoniumlike $Z_b$ states
and the bottom meson loops in the heavy quark spin flip transitions
$\Upsilon(4S) \to h_b(1P,2P) \pi^+\pi^-$.
Depending on the constructive or destructive interferences between
the $Z_b$-exchange and the bottom meson loops mechanisms, we predict two possible branching ratios for each process:
BR$_{\Upsilon(4S) \to h_b(1P)\pi^+\pi^-}\simeq\big(1.2^{+0.8}_{-0.4}\times10^{-6}\big)$ or $\big( 0.5^{+0.5}_{-0.2}\times10^{-6}\big)$, and
BR$_{\Upsilon(4S) \to h_b(2P)\pi^+\pi^-}\simeq \big(7.1^{+1.7}_{-1.1}\times10^{-10}\big)$ or $\big(    2.4^{+0.2}_{-0.1}\times10^{-10}\big)$.
The bottom meson loops contribution is found to be much larger than the $Z_b$-exchange contribution
in the $\Upsilon(4S) \to h_b(1P) \pi\pi$ transitions, while it can not produce decay rates comparable to
the heavy quark spin conserved  $\Upsilon(4S) \to \Upsilon(1S,2S) \pi\pi$ processes. We also predict the branch fractions of
$\psi(3S,4S) \to h_c(1P)\pi^+\pi^-$ contributed from the charm meson loops.

\end{abstract}

\maketitle

\newpage
\section{Introduction}

The hadronic transitions $\Upsilon(mS) \to
\Upsilon(lS) \pi \pi $ and $\Upsilon(mS) \to
h_b(nP) \pi \pi $ are important processes for understanding the heavy-quarkonium dynamics and
low-energy QCD. Because the bottomonia
are expected to be compact and nonrelativistic, the method of the
QCD multipole
expansion (QCDME)~\cite{Voloshin1980,Novikov1981,Kuang1981,Kuang2006} is
often used in analysis of these transitions, where the pions emitted come
from the hadronization of soft gluons. The decay rates of $\Upsilon(2S,3S) \to \Upsilon(1S,2S) \pi \pi $ can be well described by the
method of QCDME~\cite{ref:Kuang}.
Since the total spin of $b\bar{b}$ system in $\Upsilon(mS)$ and $h_b(nP)$ are 1 and
0, respectively, and thus in general the heavy quark spin flip $\Upsilon(mS) \to h_b(nP)\pi\pi$ processes
are expected to be suppressed compared with the heavy quark spin conserved $\Upsilon(mS) \to \Upsilon(nS)\pi\pi$ processes.
Within the framework of QCDME, references~\cite{ref:standardmultipoleTuan,ref:Kuang,ref:kuangyan}
predict that the branching fraction of $\Upsilon(3S) \to \Upsilon(1P) \pi \pi$ is suppressed by 2 orders of magnitude relative to that of
$\Upsilon(3S) \to \Upsilon(1S) \pi \pi$, while Ref.~\cite{ref:Voloshin} predicts an at least 3 orders of magnitude suppression. The prediction of Ref.~\cite{ref:Voloshin} is supported by the experimental data~\cite{Lees:2011bv}.
While in the decay processes $\Upsilon(5S)
\rightarrow \Upsilon(lS) \pi^+ \pi^- $ ($l=1, 2, 3$) and
$\Upsilon(5S) \rightarrow h_b(nP) \pi^+ \pi^- $ ($n=1,
2$) where the two charged bottomoniumlike resonances $Z_b(10610)^\pm$ and
$Z_b(10650)^\pm$ were observed, the $\Upsilon(5S) \rightarrow h_b(nP) \pi^+ \pi^- $
proceed at a rate comparable to the $\Upsilon(5S)
\rightarrow \Upsilon(lS) \pi^+ \pi^- $ processes~\cite{Belle2011:1,Belle2012:1}.
The mechanism that mitigates the expected suppression has remained controversial. In Refs.~\cite{Chen:2011pv,Swanson:2014tra}, the $\Upsilon(5S) \rightarrow h_b(nP) \pi^+ \pi^- $ processes are interpreted
via bottom meson loops mechanism, while genuine $S$-matrix $Z_b$ poles are required as in Refs.~\cite{Guo:2014iya,Huo:2015uka,Guo:2016bjq,Wang:2018jlv}. Note that the meson loops mechanism has been explored by many previous works~\cite{Chen:2011qx,Chen:2011zv,Meng:2007tk,Meng:2008bq,Wang:2016qmz,Simonov:2008qy,Wang:2015xsa} to study the dipion and $\eta$ transitions of higher charmonia and bottomonia since the branch ratios and the dipion invariant mass spectra cannot be described by using QCDME.

In this work, we will study that whether the bottom meson loops mechanism can produce the $\Upsilon(4S) \rightarrow h_b(nP) \pi^+ \pi^-$ transitions at the decay ratios comparable with $\Upsilon(4S) \rightarrow \Upsilon(lS) \pi^+ \pi^-$.
Since in the dipion emission processes of the $\Upsilon(4S)$ the crossed-channel exchanged $Z_b$ can not be on-shell, one may expect that these transitons are good channels to study the bottom meson loops' effect.
In our previous works~\cite{Chen:2016mjn,Chen:2019gty}, using the nonrelativistic effective field theory (NREFT) we calculated the effects
of the bottom meson
loops as well as the $Z_b$-exchange in the $\Upsilon(4S) \to
\Upsilon(1S,2S) \pi \pi $ processes, and found that the experimental data can be described well.
Here within the same theoretical scheme, we will calculate the contributions of the bottom meson loops
and the $Z_b$-exchange in the $\Upsilon(4S) \rightarrow h_b(nP) \pi^+ \pi^-$ processes, and give the theoretical predictions of the decay branching ratios. We find that the contribution of the bottom meson loops is much larger than that of the $Z_b$-exchange in the $\Upsilon(4S) \rightarrow h_b(1P) \pi^+ \pi^-$ process, while it
can not produce a rate comparable with $\Upsilon(4S) \rightarrow \Upsilon(1S,2S) \pi^+ \pi^-$.

This paper is organized as follows. In Sec.~\ref{theor}, the theoretical framework is described in detail.
In Sec.~\ref{pheno}, we give the theoretical predictions for the decay branching fractions of $\Upsilon(4S) \rightarrow h_b(1P,2P) \pi^+ \pi^-$, and discuss the contributions of different mechanisms. A summary will be
given in Sec.~\ref{conclu}.

\section{Theoretical framework}\label{theor}
\subsection{Lagrangians}

To calculate the contribution of the mechanism $\Upsilon(mS) \to
Z_b\pi \to h_b(nP) \pi\pi$, we need the effective Lagrangians for the $Z_b \Upsilon\pi$ interaction and $Z_b h_b\pi$ interaction~\cite{Guo2011},
\be\label{LagrangianZbUppi} \L_{Z_b\Upsilon\pi} =
\sum_{j=1,2}C_{Z_{bj}\Upsilon(mS)\pi} \Upsilon^i(mS) \bra
{Z^i_{bj}}^\dagger u_\mu \ket v^\mu +\mathrm{h.c.} \,,
\ee
\be\label{LagrangianZbhbpi} \L_{Z_b h_b\pi} =
\sum_{j=1,2}g_{Z_{bj}h_b(nP)\pi}\epsilon_{ijk}\bra {Z^i_{bj}}^\dagger u^j \ket h_b^k +\mathrm{h.c.} \,,
\ee
where
$Z_{b1}$ and $Z_{b2}$ denote $Z_b(10610)$ and
$Z_b(10650)$, respectively, and $v^\mu=(1,\vec{0})$ is the velocity of the heavy quark. The $Z_b$ states are collected in the
matrix as
\begin{equation} \label{eq:Z-field}
Z^i_{bj}=
 \left( {\begin{array}{*{2}c}
   \frac{1}{\sqrt{2}}Z^{0i}_{bj} & Z^{+i}_{bj}   \\
   Z^{-i}_{bj} & -\frac{1}{\sqrt{2}}Z^{0i}_{bj}
\end{array}} \right)\,.
\end{equation}
The pions as Goldstone bosons of the spontaneous breaking of the chiral symmetry can be parametrized as
\begin{align}
u_\mu &= i \left( u^\dagger\partial_\mu u\, -\, u \partial_\mu u^\dagger\right) \,, \qquad
u = \exp \Big( \frac{i\Phi}{\sqrt{2}F_\pi} \Big)\,, \nonumber\\
\Phi &=
 \begin{pmatrix}
   {\frac{1}{\sqrt{2}}\pi ^0 } & {\pi^+ }   \\
   {\pi^- } & {-\frac{1}{\sqrt{2}}\pi ^0}  \\
 \end{pmatrix} , \label{eq:u-phi-def}
\end{align}
where $F_\pi=92.2\MeV$ is the pion decay constant.

To calculate the box diagrams, we need the Lagrangian for the
coupling of the $\Upsilon$ to the bottom mesons and the coupling of the $h_b$ to the bottom mesons~\cite{Guo2009:PRL,Guo2011},
\begin{equation}\label{LagrangianJHH}
\L_{\Upsilon HH}=\frac{i\, g_{JHH}}{2}\bra J^\dag H_a
\boldsymbol{\sigma}\cdot \!\overleftrightarrow{\partial}\!
\bar{H}_a\ket + {\rm h.c.}\,,
\end{equation}
\begin{equation}\label{LagrangianhbHH}
\L_{h_b HH}=\frac{i\, g_{1}}{2}\bra h_b^{\dag i} H_a
\sigma^i\bar{H}_a\ket + {\rm h.c.}\,,
\end{equation}
where $J \equiv
\vec{\Upsilon} \cdot \boldsymbol{\sigma}+\eta_b$ denotes the heavy quarkonia spin multiplet,
$H_a=\vec{V}_a \cdot \boldsymbol{\sigma}+P_a$ with
$P_a(V_a)=(B^{(*)-},\bar{B}^{(*)0})$ collects the bottom mesons, and
$A\overleftrightarrow{\partial}\!B\equiv
A(\overrightarrow{\partial}B)-(\overrightarrow{\partial}A)B$.
We also need the Lagrangian for the axial coupling of the pion fields to the bottom and antibottom mesons, which at the lowest order
in heavy-flavor chiral perturbation theory is given
by~\cite{Burdman:1992gh,Wise:1992hn,Yan:1992gz,Casalbuoni:1996pg,Mehen2008}
\begin{equation}\label{LagrangianHHPhi}
\L_{HH\Phi}= \frac{g_\pi}{2} \bra \bar{H}_a^\dagger  \boldsymbol{\sigma} \cdot \vec{u}_{ab}
\bar{H}_b\ket -\frac{g_\pi}{2} \bra H_a^\dagger H_b
\boldsymbol{\sigma} \cdot \vec{u}_{ba} \ket,
\end{equation}
where $u^i=-\sqrt{2}\partial^i \Phi/F + \mathcal{O}(\Phi^3)$ denotes the three-vector components of $u_\mu$
as defined in Eq.~\eqref{eq:u-phi-def}.
Here we will use $g_\pi=0.492\pm0.029$ from a recent lattice QCD
calculation~\cite{Bernardoni:2014kla}.

\subsection{Power counting of the loops}\label{subsection:powercounting}

\begin{figure}
\centering
\includegraphics[width=\linewidth]{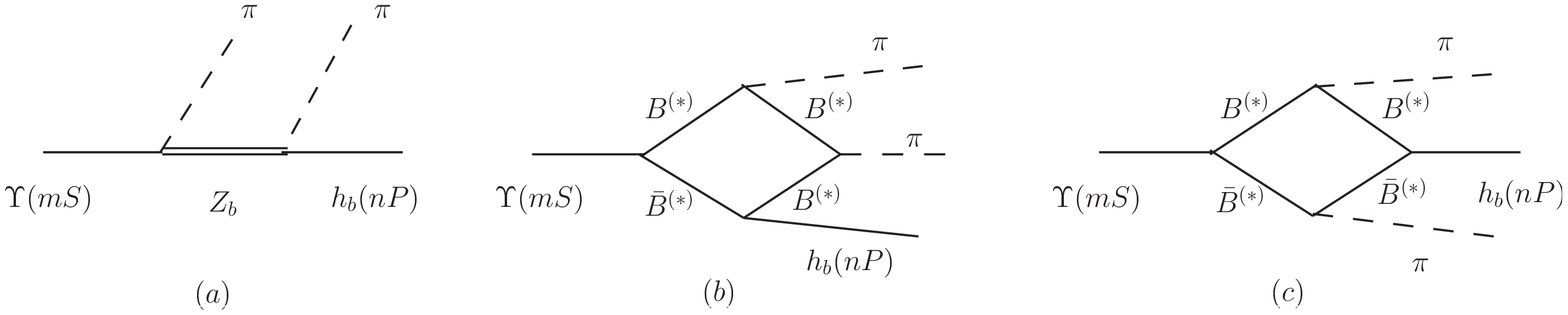}
\caption{Feynman diagrams considered for the $\Upsilon(mS)
\rightarrow h_b(nP) \pi \pi $ processes. The crossed diagrams
of (a) and (b) are not shown explicitly.
}\label{fig.FeynmanDiagram}
\end{figure}
\begin{figure}
\centering
\includegraphics[width=\linewidth]{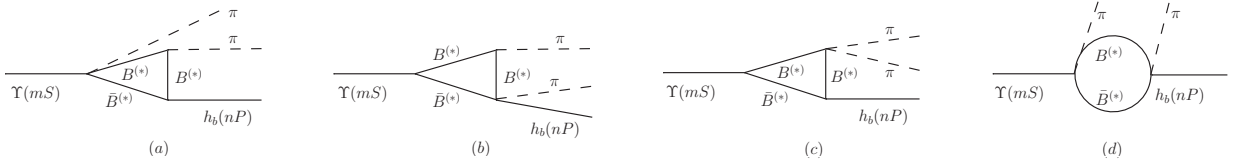}
\caption{The loop diagrams not considered in the calculations.
The corresponding power counting arguments are given in the main text.}
\label{fig.FeynmanDiagramTriangle}
\end{figure}

Since the $\Upsilon(4S)$ meson is above the $B\bar{B}$ threshold and
decays predominantly into $B\bar{B}$ pairs, the loop mechanism with
intermediate bottom mesons may be important in the
transitions $\Upsilon(4S) \rightarrow h_b(nP)
\pi^+\pi^-$. Following the formalism set-up based on NREFT~\cite{Guo2009:PRL,Guo2011:effect,Martin2013},
we will analyze the power counting of different kinds of loops. In NREFT, the expansion parameter is the velocity of the intermediate
heavy meson, namely
$\nu_X=\sqrt{|m_X-m_{B^{(*)}}-m_{B^{(*)}}|/m_{B^{(*)}}}$, which is small since the bottomonia $X$ are close to the
$B^{(*)}\bar{B}^{(*)}$ thresholds. In this power counting, each nonrelativistic propagator scale as $1/\nu^2$,
and the measure of one-loop integration scales as $\int d^4 l \sim \nu^5$.

There are five different kinds of loop
contributions, namely the box diagrams displayed in
Fig.~\ref{fig.FeynmanDiagram}\,(b), (c), the triangle diagrams
displayed in Fig.~\ref{fig.FeynmanDiagramTriangle}\,(a)--(c) and the bubble loop in Fig.~\ref{fig.FeynmanDiagramTriangle}\,(d). We analyze them one by one as follows:

First we analyze the power counting of the box diagrams, namely Fig.~\ref{fig.FeynmanDiagram}\,(b), (c).
As indicated in Eq.~\eqref{LagrangianHHPhi}, the vertex of $B^{(*)}B^{(*)}\pi$ is proportional to the
external momentum of the pion $q_\pi$. The $\Upsilon B^{(*)}\bar{B}^{(*)}$ vertex is in a $P$-wave, and the
$h_bB^{(*)}\bar{B}^{(*)}$ vertex is in an $S$-wave, so the loop momentum must contract with
the external pion momentum and hence the $P$-wave vertex scales
as $\mathcal{O}(q_\pi)$. Thus the box diagrams scales as $\nu^5 q_\pi^3/\nu^8= q_\pi^3/\nu^3.$

As for the triangle diagram Fig.~\ref{fig.FeynmanDiagramTriangle}\,(a), the leading
$\Upsilon B^{(*)}\bar{B}^{(*)}\pi$ vertex given by $g_{JHH\pi}\bra J \bar{H}_a^\dag
H_b^\dag\ket u_{ab}^0$~\cite{Mehen2013}
is proportional to the energy of the pion, $E_\pi\sim q_\pi$.
Therefore, Fig.~\ref{fig.FeynmanDiagramTriangle}\,(a) is counted as
$ m_B\nu^5 q_\pi^2/\nu^6 =  m_B q_\pi^2/\nu$, where the factor $m_B$ has been
introduced to match the dimension with the scaling for the box diagrams.

In Fig.~\ref{fig.FeynmanDiagramTriangle}\,(b), the leading $h_b B^{(*)}\bar{B}^{(*)}\pi$ vertex given by $g_{h_b HH\pi}\bra h_b^{\dag i} H_a
\sigma^j \bar{H}_b\ket \epsilon_{ijk} u_{ab}^k$~\cite{Mehen2008}
is proportional to the momentum the pion $q_\pi$. The loop momentum due to the $\Upsilon B^{(*)}\bar{B}^{(*)}$ coupling has to contract with the external pion momentum.
Thus, Fig.~\ref{fig.FeynmanDiagramTriangle}\,(b) scales as
$ \nu^5  q_\pi^3/\nu^6= q_\pi^3/\nu$.

The leading
$B^{(*)}B^{(*)}\pi\pi$ vertex comes from the chiral
derivative term $\bra H_a^\dag(iD_0)_{ba}H_b\ket =\bra
H_a^\dag(i\partial_0-iV_0)_{ba}H_b\ket$~\cite{Stewart,Mehen2006},
in which the pion pair produced by the vector current,
$V^\mu=\frac{1}{2}(u^\dag \partial^\mu u+u\partial^\mu u^\dag)$, cannot form a
positive-parity and $C$-parity state, therefore this leading vertex does not contribute to the
$\Upsilon(mS) \rightarrow h_b(nP) \pi \pi $ processes.
Isoscalar, $PC=++$ pion pairs only enter in the next order
$\mathcal{O}(q_\pi^2)$ from point vertices. Therefore, Fig.~\ref{fig.FeynmanDiagramTriangle}\,(c) scales as $\nu^5 q_\pi^3  /\nu^6= q_\pi^3/\nu$.

In Fig.~\ref{fig.FeynmanDiagramTriangle}\,(d), both the initial vertex and the final vertex are proportional to $q_\pi$, so the bubble loop scales as $m_B\nu^5 q_\pi^2 /\nu^4= m_B q_\pi^2\nu$.

Therefore, we expect that the ratios of the contributions of the box diagrams, triangle diagram Fig.~\ref{fig.FeynmanDiagramTriangle}\,(a)--(c),
and the bubble loop Fig.~\ref{fig.FeynmanDiagramTriangle}\,(d) are
\bqa \label{eq.Ratios}
&&\frac{ q_\pi^3}{\nu^3} : \frac{ m_B q_\pi^2}{\nu} : \frac{ q_\pi^3}{\nu} : \frac{ q_\pi^3}{\nu}
:  m_B q_\pi^2\nu \nn\\
= && 1 : \frac{m_B \nu^2}{q_\pi} :  \nu^2 :  \nu^2
: \frac{m_B \nu^4}{q_\pi}\,,
\eqa
where $q_\pi \simeq (m_{\Upsilon(4S)}-m_{h_b(nP)})/2$ and $\nu=(\nu_{\Upsilon(4S)}+\nu_{h_b(nP)})/2$, with $\nu_{\Upsilon(4S)}\simeq 0.06$,
$\nu_{h_b(1P)}\simeq 0.35$, and $\nu_{h_b(2P)}\simeq 0.24$.
Thus for the $\Upsilon(4S) \to h_b(1P) \pi^+\pi^-$ transition, the ratios in Eq.~\eqref{eq.Ratios}
are $1: 0.67 : 0.04 : 0.04 : 0.03$.
For the $\Upsilon(4S) \to h_b(2P) \pi^+\pi^-$ transition, the ratios are $1: 0.75 : 0.02 : 0.02 : 0.02$.
Therefore according to the power counting the box diagrams and the triangle diagram in Fig.~\ref{fig.FeynmanDiagramTriangle}\,(a)
are dominant among the loop contributions, and they are of the same order. While the $\Upsilon(4S)$ is below the $B^{(*)}\bar{B}^{(*)}\pi$ threshold and the coupling $g_{JHH\pi}$ in the triangle diagram Fig.~\ref{fig.FeynmanDiagramTriangle}\,(a) is unknown. Thus for a rough estimation of the loop contributions, we will only calculate the box diagrams in the present study.
Note that all the box and triangle loop contributions discussed here are ultraviolet-finite,
and do not require the additional introduction of counterterms.

\subsection{Tree-level amplitudes and box diagram calculation}\label{AmplitudesCalculation}

The decay amplitude for
\be
\Upsilon(mS)(p_a) \to h_b(nP)(p_b) \pi(p_c)\pi(p_d)
\ee
is described in terms of the Mandelstam variables
\begin{align}
s &= (p_c+p_d)^2 , \qquad
t=(p_a-p_c)^2\,, \qquad u=(p_a-p_d)^2 \,.
\end{align}

Using the effective Lagrangians in Eqs.~\eqref{LagrangianZbUppi} and~\eqref{LagrangianZbhbpi},
the tree amplitude of $\Upsilon(mS) \to Z_b\pi \to h_b(nP)\pi\pi$ can be obtained
\begin{align}\label{eq.MZb}
M_{Zb}=\frac{2
\sqrt{m_{\Upsilon(mS)}m_{h_b(nP)}}}{F_\pi^2}\epsilon_{abj}\epsilon_{\Upsilon(mS)}^a\epsilon_{h_b(nP)}^b\sum_{i=1,2}m_{Z_{bi}} C_{Z_{bi}\Upsilon(mS)\pi}g_{Z_{bi}h_b(nP)\pi}
\Big\{ p_c^0 p_d^j\frac{1}{t-m_{Z_{bi}}^2}+p_d^0 p_c^j\frac{1}{u-m_{Z_{bi}}^2} \Big\}\,.
\end{align}
Notice that the nonrelativistic normalization factor $\sqrt{m_Y}$ has been multiplied to the amplitude for every heavy particle, with
$Y=\Upsilon(mS), h_b(nP), Z_{bi}$. The widths of the $Z_b$ states are neglected in the present study, since they are of the
order of $10\MeV$ and are much smaller than the difference between the $Z_b$ masses and the $\Upsilon(mS)\pi/h_b(nP)\pi$ threshold.%

Now we discuss the calculation of the box diagrams. In the box diagrams
Fig.~\ref{fig.FeynmanDiagram}\,(b) and (c), we denote the top left
intermediate bottom meson as $M1$, and the other intermediate bottom mesons
as $M2$, $M3$, and $M4$, in counterclockwise order.
Concerned with the pseudoscalar or vector content of $[M1,M2,M3,M4]$,
there are twelve possible patterns and we number them in order:
1, $[PPPV]$; 2, $[PPVV]$; 3, $[PVPV]$; 4, $[PVVP]$; 5, $[VVPP]$; 6, $[VPVP]$;
7, $[VPPV]$; 8, $[PVVV]$; 9, $[VPVV]$; 10, $[VVPV]$; 11, $[VVVP]$; 12, $[VVVV]$.
For each pattern, we also need to consider six possibilities of different flavor of the intermediate
bottom mesons: $[B^{(*)+},B^{(*)-},B^{(*)+},B^{(*)0}]$, $[B^{(*)+},B^{(*)-},{\bar B}^{(*)0},B^{(*)0}]$,
$[B^{(*)0},{\bar B}^{(*)0},B^{(*)0},B^{(*)+}]$, $[B^{(*)-},B^{(*)+},B^{(*)-},{\bar B}^{(*)0}]$,
$[{\bar B}^{(*)0},B^{(*)0}, B^{(*)+},B^{(*)-}]$, and  $[{\bar B}^{(*)0},B^{(*)0},{\bar B}^{(*)0},B^{(*)-}]$.
The full amplitude contains the sum of all possible ones.

For the tensor reduction of the loop integrals it is convenient to
define $\vec q = -\vec p_b$ and the perpendicular momentum $\vec q_\perp=\vec p_c-\vec
q(\vec q\cdot \vec p_c)/\vec q^2$, which satisfy $\vec q \cdot \vec q_\perp=0$.
The result of the amplitude of the box diagrams can be written as
\begin{align}\label{eq.Mloop}
M_{\text{loop}}=\epsilon_{\Upsilon(mS)}^a\epsilon_{h_b(nP)}^b
\Big\{ \epsilon_{abi}q^i A_1+ \epsilon_{abi}q_\perp^i A_2 +\epsilon_{bij}q^i q_\perp^j q_\perp^a A_3
+\epsilon_{bij}q^i q_\perp^j q^a A_4 +\epsilon_{aij}q^i q_\perp^j q_\perp^b A_5 +\epsilon_{aij}q^i q_\perp^j q^b A_6 \Big\}\,.
\end{align}
Details on the analytic calculation
of the box diagrams and the explicit expressions of $A_i$ $(i=1,2,...,6)$ are given in Appendix~\ref{app:box}.

The decay width for $\Upsilon(mS) \to h_b(nP)\pi\pi$ is given by
\bqa
\Gamma=\int_{s_-}^{s_+} \int_{t_-}^{t_+}\frac{|M_{Zb}+M_{\text{loop}}|^2 ds dt}{768 \pi^3 m_{\Upsilon(mS)}^3}\,,
\eqa
where the lower and upper limits are given as
\bqa
s_-&=&4m_\pi^2,\nonumber\\
s_+&=&(m_{\Upsilon(mS)}-m_{h_b(nP)})^2,\nonumber\\
t_{\pm}&=&\frac{1}{4s}\Big\{ (m_{\Upsilon(mS)}^2-m_{h_b(nP)}^2)^2-\big[\lambda^{\frac{1}{2}}(s,m_\pi^2,m_\pi^2)\mp
\lambda^{\frac{1}{2}}(m_{\Upsilon(mS)}^2,s,m_{h_b(nP)}^2)\big]^2\Big\},\nonumber\\
\lambda(a,b,c)&=& a^2+b^2+c^2-2(ab+ac+bc)\,.
\eqa

\section{Phenomenological discussion}\label{pheno}

To estimate the contribution of the $Z_b$-exchange mechanism we need to know the coupling strengths of
$Z_{b} \Upsilon(4S)\pi$ and $Z_{b} h_b(nP)\pi$.
The mass difference between $Z_b(10610)$ and $Z_b(10650)$
is much smaller than the difference between their masses and the $\Upsilon(mS)\pi/h_b(nP)\pi$ threshold, and they have the same quantum numbers and thus the same
coupling structures as dictated by Eqs.~\eqref{LagrangianZbUppi} and~\eqref{LagrangianZbhbpi}. Therefore it is very difficult to distinguish their effects from each other in the dipion transitions of $\Upsilon(4S)$, so we only use
one $Z_b$, the $Z_b(10610)$, which approximately combine both $Z_b$ states' effects. In Ref.~\cite{Chen:2019gty}, we has studied the $\Upsilon(4S) \to \Upsilon(mS)\pi\pi$ processes
to extract the coupling
constant $|C_{Z_{b}\Upsilon(4S)\pi}|=(3.3 \pm 0.1)\times 10^{-3}$,  which containing effects from both $Z_b$ states.
For the couplings of $Z_{b} h_b(nP)\pi$, in principle they can be extracted from
the partial widths of the $Z_b$ states decay into $h_b(nP)\pi\,(n=1,2)$
\begin{equation}\label{eq.gZbhbpi}
|g_{Z_{b}h_b\pi}|=\Bigg\{\frac{6\pi F_\pi^2 m_{Z_b} \Gamma_{Z_b \to
h_b\pi}}{|\vec{p}_f|^3 m_{h_b}}
\Bigg\}^{\frac{1}{2}} \,,
\end{equation}
where $|\vec{p}_f|\equiv
\lambda^{1/2}\big(m_{Z_b}^2,m_{h_b}^2,m_\pi^2\big)/(2m_{Z_b})$.
The branching fractions of the decays
of both $Z_b$ states into $h_b(nP)\pi\,(n=1,2)$ have been given in~\cite{Garmash:2015rfd}, where the $Z_b$ line shapes were described using Breit-Wigner forms.
If we naively use these branching fractions, we would obtain
\bqa \label{eq.gZbhbpiNaive}
|g_{Z_{b1}h_b(1P)\pi}^\text{naive}|&=&0.019 \pm 0.003\,,\nn\\
|g_{Z_{b2}h_b(1P)\pi}^\text{naive}|&=&0.021 \pm 0.003\,,\nn\\
|g_{Z_{b1}h_b(2P)\pi}^\text{naive}|&=&0.068 \pm 0.011\,,\nn\\
|g_{Z_{b2}h_b(2P)\pi}^\text{naive}|&=&0.077 \pm 0.010\,.
\eqa
Here all the $Z_{b} h_b\pi$ couplings are labeled by a superscript ``naive'' since this is not the appropriate way to extract the coupling strengths in this case; the $Z_b$
states are very close to the $B^{(*)}\bar B^*$ thresholds, and thus the Flatt\'e parametrization for the $Z_b$ spectral functions should be used, which will lead to much larger partial
widths into $(b\bar{b}\pi)$ channels, and thus the relevant coupling
strengths. As analyzed in Ref.~\cite{Chen2016}, in the the Flatt\'e parametrization the sum of the partial
widths of the $Z_b(10610)$ other than that for the $B\bar B^*$ channel should be
larger than the nominal width, which is about $20\MeV$. While summing over all the
$\Upsilon(nS)\pi\,(n=1,2,3)$ and $h_b(mP)\pi\,(m=1,2)$ branching fractions in
Ref.~\cite{Garmash:2015rfd} gives about 14\% or $3\MeV$ in terms of partial widths. Therefor for a rough estimation we will use three times the results from Eq.~\eqref{eq.gZbhbpiNaive},
namely
\bqa \label{eq.gZbhbpiValues}
|g_{Z_{b}h_b(1P)\pi}|&\simeq&0.057 \,,\nn\\
|g_{Z_{b}h_b(2P)\pi}|&\simeq&0.204 \,.
\eqa
We will find that even considering this enlarging factor of three for the couplings $|g_{Z_{b}h_b(nP)\pi}|$, the $Z_b$-exchange contribution is still
much smaller than the bottom meson loops contributions.

In the calculation of the box diagrams, the coupling strength $g_{JHH(4S)}$ can be extracted from the measured
open-bottom decay widths of the $\Upsilon(4S)$, and we have $g_{JHH(4S)}=1.43\pm 0.01\GeV^{-3/2}$. For the $h_b B^{*}\bar{B}^{*}$ coupling $g_1$,
we can use the results from Ref.~\cite{Guo2011}.
In~\cite{Guo2011}, the $Z_b$-exchange mechanism in the $\Upsilon(5S) \to h_b(1P,2P)\pi\pi$ processes has been studied assuming the $Z_b$ states are $B^{(\ast)}\bar{B}^{\ast}$ bound states, and the physical coupling of the $Z_b$ states to the bottom and anti-bottom mesons, $z_1$, as well as the product $g_1z_1$
have been determined.\footnote{In~\cite{Guo2011}, in order to reduce the number of free parameters, the couplings of $h_b(1P) B^{*}\bar{B}^{*}$ and $h_b(2P) B^{*}\bar{B}^{*}$ are assumed to be the same.} Using their results $z_1=0.75^{+0.08}_{-0.11} \GeV^{-1/2}$
and $g_1 z_1=0.40\pm 0.06 \GeV^{-1}$, we can extract that $g_1=0.53^{+0.19}_{-0.13} \GeV^{-1/2}$.

Using the coupling strengths above, we can predict the decay branching fractions of $\Upsilon(4S) \to h_b(1P,2P)\pi^+\pi^-$.
Depending on the sign of the couplings in Eq.~\eqref{eq.gZbhbpiValues}, the interferences can be constructive or destructive between
the $Z_b$-exchange and box graph mechanisms, so there are two possible results for each process
\bqa  \label{eq.BR4Sto1P2Pcase1}
\text{BR}_{\Upsilon(4S) \to h_b(1P)\pi^+\pi^-}&\simeq& \big(1.2^{+0.8}_{-0.4}\times10^{-6}\big) \quad \text{or} \quad \big( 0.5^{+0.5}_{-0.2}\times10^{-6}\big) \,,\nn\\
\text{BR}_{\Upsilon(4S) \to h_b(2P)\pi^+\pi^-}&\simeq& \big(7.1^{+1.7}_{-1.1}\times10^{-10}\big) \quad \text{or} \quad\big(    2.4^{+0.2}_{-0.1}\times10^{-10}\big)\,.
\eqa
We find that the $\text{BR}_{\Upsilon(4S) \to h_b(1P)\pi^+\pi^-}$ is at least one order of magnitude smaller than the branching fractions
$\text{BR}_{\Upsilon(4S) \to \Upsilon(1S,2S)\pi^+\pi^-}$, which are about $8\times10^{-5}$ given in PDG~\cite{PDG}, and the $\text{BR}_{\Upsilon(4S) \to h_b(2P)\pi^+\pi^-}$
is tiny due to the very small phase space.
We discuss the $\Upsilon(4S) \to h_b(1P)\pi^+\pi^-$
transition in more details. To illustrate the effects of the $Z_b$-exchange and box graph mechanisms in $\Upsilon(4S)\to h_b(1P)\pi\pi$, we give the predictions only including the $Z_b$-exchange terms or
only including the box diagrams
\bqa  \label{eq.BR4Sto1PDifferentParts}
\text{BR}_{\Upsilon(4S) \to h_b(1P)\pi^+\pi^-}^{Z_b}&=&0.6^{+0.1}_{-0.1}\times10^{-7}\,,\nn\\
\text{BR}_{\Upsilon(4S) \to h_b(1P)\pi^+\pi^-}^{\text{Box}}&=&0.8^{+0.7}_{-0.3}\times10^{-6}\,.
\eqa
One observes that the bottom meson loops contribution is much larger than the $Z_b$-exchange contribution, while it is 2 orders of magnitude smaller than the $\Upsilon(4S) \to \Upsilon(1S,2S)\pi^+\pi^-$ transitions.
Note that there is no calculation of the direct gluon hadronization mechanism contribution within QCDME for
the $\Upsilon(4S) \to h_b(1P)\pi^+\pi^-$ process in the literature. In references~\cite{ref:standardmultipoleTuan,ref:Kuang,ref:kuangyan,ref:Voloshin}, QCDME predict that the branching fraction of $\Upsilon(3S) \to h_b(1P) \pi \pi$ is 2-3 orders of magnitude suppressed relative to that of
$\Upsilon(3S) \to \Upsilon(1S) \pi \pi$, and the 3 orders of magnitude suppression is supported by the experiment~\cite{Lees:2011bv}.
Since the mass difference between the $\Upsilon(4S)$ and $h_b(1P)$ is about 0.68 GeV, the pions in the $\Upsilon(4S) \to h_b(1P)\pi^+\pi^-$
process also can be considered to be in the soft region. If one approximates that the gluon hadronization mechanism within QCDME in $\Upsilon(4S) \to h_b(1P)\pi^+\pi^-$ is also 2-3 orders of magnitudes suppressed relative to that in the $\Upsilon(4S) \to \Upsilon(1S)\pi^+\pi^-$ process, as in the $\Upsilon(3S)$ decay cases, then one obtains that the gluon hadronization mechanism contribution is at most at the same order of the bottom meson loops contribution.
Due to the lack of exact information concerning the gluon hadronization within QCDME and the neglecting of triangle diagram Fig.~\ref{fig.FeynmanDiagramTriangle}\,(a) as discussed in section~\ref{subsection:powercounting}, it is important to keep in mind that the results presented in this paper are order-of-magnitude estimates.

\begin{figure}
\centering
\includegraphics[width=\linewidth]{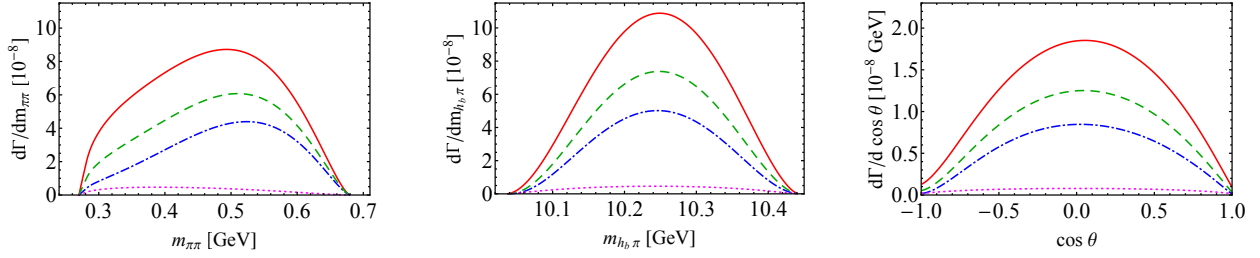}
\caption{Theoretical predictions of the distributions of the $\pi\pi$ and $h_b\pi$ invariant mass spectra, and the helicity angular distributions in the $\Upsilon(4S) \to h_b(1P) \pi \pi$ process.
The darker green dashed, magenta dotted, red solid, and blue dot-dashed lines represent the contributions of the box diagrams, $Z_b$-exchange, the sum of them with constructive interference, and the sum with destructive interference, respectively.
}
\label{fig.distribution}
\end{figure}

In Fig.~\ref{fig.distribution}, we plot the distributions of the $\pi\pi$ and $h_b\pi$ invariant mass spectra, and the distribution of cos$\theta$, where $\theta$ is defined as the angle between the initial $\Upsilon(mS)$ and the $\pi^+$ in the rest frame of the $\pi\pi$
system. To illustrate the effects of different mechanisms, the contributions of the box diagrams, $Z_b$-exchange, the sum of them with constructive interference, and the sum with destructive interference are shown as the darker green dashed, magenta dotted, red solid, and blue dot-dashed lines, respectively. One observes that there is a broad bump around 0.5 GeV in the dipion invariant mass distribution.
While the $\pi\pi$ invariant mass spectra with unknown normalization predicted within QCDME in Ref.~\cite{ref:Kuang} display a peak at low $\pi\pi$ masses. Thus the $\pi\pi$ invariant mass spectra can be useful to identify the effects of bottom meson loops and the gluon hadronization mechanism with future experimental data. Also one observes that the angular distribution is far from flat. In the $\Upsilon(mS) \to h_b(nP) \pi \pi$ process, isospin conservation combined with Bose symmetry requires the pions to have even relative angular momentum. Therefore it means that there is a large $D$-wave component from the box diagrams, if neglecting yet higher partial waves.

The $\Upsilon(mS) \to h_b(nP)\pi\pi$ are heavy quark spin flip processes and they are forbidden in
the heavy quark limit.
We have checked that in the heavy quark limit, \textit{i.e.}\
$m_B=m_{B^*}$, all the box diagrams are cancelled with each other so the bottomed loops contribute nothing
to the $\Upsilon(mS)\to h_b(nP)\pi\pi$ transitions. With the small mass splitting of $B$ and $B^*$ in the real world, in Eqs.~\eqref{eq.BR4Sto1P2Pcase1} and~\eqref{eq.BR4Sto1PDifferentParts} one observes that the bottomed meson loops contribution does not produce $\Upsilon(4S)\to h_b(1P)\pi\pi$ at a rate comparable to the heavy quark spin conserved $\Upsilon(4S)\to \Upsilon(1S,2S)\pi\pi$ transitions.
Note that the datasets collected at $\Upsilon(4S)$ by BABAR and Belle II collaborations are $471\times 10^6$ and $772\times 10^6$ ~\cite{Kou:2018nap}, respectively, and thus they should contain several hundreds of $\Upsilon(4S)\to h_b(1P)\pi\pi$ events according to our calculation. We hope future experimental analysis by BABAR and Belle can test our predictions.
On the other hand, as stated in the introduction, the observed $\Upsilon(5S) \rightarrow h_b(nP) \pi^+ \pi^- $
proceed at a rate comparable to the $\Upsilon(5S)
\rightarrow \Upsilon(lS) \pi^+ \pi^- $ processes~\cite{Belle2011:1,Belle2012:1}. The enhancements may come from the effects of the on-shell $Z_b$ exchange and the two-cut condition complexity of the bottom meson loops in the $\Upsilon(5S)$ decays. A detailed analysis of the $\Upsilon(5S) \rightarrow h_b(nP) \pi^+ \pi^- $ processes is beyond the scope of this paper.

Due to the similarity between the bottomonium and charmonium families, we can extend the box diagrams calculation to give a rough estimation of the branch fractions of the $\psi(3S) (\psi(4040)) \to h_c(1P)\pi^+\pi^-$ and $\psi(4S) (\psi(4415)) \to h_c(1P)\pi^+\pi^-$ transitions. The relevant Feynman diagrams can be obtained by replacing the external $\Upsilon(mS)$ and $h_b(nP)$ by $\psi(mS)$ and $h_c(nP)$, respectively, and replacing the intermediate $B^{(*)}$ by $D^{(*)}$ in Fig.~\ref{fig.FeynmanDiagram}\,(b) and (c). Note that the experimental decay widths of the $\psi(3S,4S) \to D^{(*)}\bar{D}^{(*)}$ transitions have not been given in PDG, and we will use the theoretical predictions of the decay widths in Ref.~\cite{Barnes:2005pb} to estimate the coupling strengths $g_{JHH(\psi(3S))}$ and $g_{JHH(\psi(4S))}$.
Since among the different decay modes $D^{(*)}\bar{D}^{(*)}$, the $D\bar{D}^{*}$ and $D^{*}\bar{D}^{*}$ modes are dominant for $\psi(3S)$ and $\psi(4S)$, respectively, we will use corresponding coupling constants in the calculation, namely $g_{JHH(\psi(3S))}=g_{\psi(3S)D\bar{D}^{*}}=0.97 \GeV^{-3/2}$ and $g_{JHH(\psi(4S))}=g_{\psi(4S)D^{*}\bar{D}^{*}}=0.25 \GeV^{-3/2}$. For the $h_c D^{*}\bar{D}^{*}$ coupling, we use the result from Ref.~\cite{Chen:2011xk}, $g_{h_c D^{*}\bar{D}^{*}}=-(\sqrt{m_{\chi_{c0}}/3})/(f_{\chi_{c0}})=-(\sqrt{3.415/3})/(0.297)  \GeV^{-1/2}=-3.59 \GeV^{-1/2}$. The predictions of the box diagrams contributions to the branch fractions of $\psi(3S,4S)  \to h_c(1P)\pi^+\pi^-$ are
\bqa  \label{eq.BRpsi3S4Stohc1Ppipi}
\text{BR}_{\psi(3S) \to h_c(1P)\pi^+\pi^-}^{\text{Box}}&=&2.9\times10^{-5}\,,\nn\\
\text{BR}_{\psi(4S) \to h_c(1P)\pi^+\pi^-}^{\text{Box}}&=&4.5\times10^{-3}\,.
\eqa
The prediction of $\text{BR}_{\psi(3S) \to h_c(1P)\pi^+\pi^-}^{\text{Box}}$ is below the upper limit given in PDG~\cite{PDG}. As expected the branch fractions $\text{BR}_{\psi(3S,4S) \to h_c(1P)\pi^+\pi^-}^{\text{Box}}$ are much larger than $\text{BR}_{\Upsilon(4S) \to h_b(1P)\pi^+\pi^-}^{\text{Box}}$, since the mass splitting of $D$ and $D^{*}$ is much larger than that of $B$ and $B^{*}$. It should be kept in
mind that this is just a preliminary rough estimation,
due to the lack of sufficient information concerning the $\psi(3S,4S) D^{(*)}\bar{D}^{(*)}$ coupling constants and the neglecting of the loop diagrams with intermediate $D_1$ state in the present calculation. A detailed theoretical study of the $\psi(3S,4S)  \to h_c(1P)\pi^+\pi^-$ transitions will be pursued in the future.

\section{Conclusions}
\label{conclu}

In this paper, we study the effects of $Z_b$ exchange and bottom meson loops
in the heavy quark spin flip transitions $\Upsilon(4S) \to h_b(nP)\pi\pi (n=1,2)$.
The bottom meson loops are treated in the NREFT scheme, in which the dominant
box diagrams are taken into account. We find that the bottom meson loops contribution is much larger than the $Z_b$-exchange contribution in the $\Upsilon(4S) \to h_b(1P)\pi\pi$ transition, while it can not produce decay rates comparable
to the heavy quark spin conserved $\Upsilon(4S) \to \Upsilon(1S,2S)\pi\pi$ processes. The theoretical
prediction of the decay rate and the dipion invariant mass spectra of $\Upsilon(4S) \to h_b(1P)\pi\pi$ in this work may be useful for identification of
the effect of the bottom meson loops with future experimental analysis.
We also predict the branch fractions of
$\psi(3S,4S) \to h_c(1P)\pi\pi$ contributed from the charm meson loops.

\section*{Acknowledgments}

We are grateful to referees' useful suggestions
and constructive remarks which are helpful in formulating the
present version of this paper.
We are grateful to Martin Cleven for the collaboration at the early stages of this study.
We acknowledge Guo-Ying Chen, Meng-Lin Du, and Qian Wang for helpful discussions, and Feng-Kun Guo for
a careful reading of the manuscript and valuable comments.
This research is supported in part by the Fundamental Research Funds
for the Central Universities under Grant
No.~FRF-BR-19-001A, and by the National Natural Science Foundation of China under Grants No.~11975028.

%
\begin{appendix}
%
\section{Remarks on the box diagrams and four-point integrals}\label{app:box}
%

In this appendix, first we will discuss the parametrization and simplification of the scalar four-point
integrals in the box diagrams. Then we introduce a tensor reduction scheme to deal with higher-rank loop integrals.
Finally, we give the amplitude of the box diagrams for the $\Upsilon(mS) \to h_b(nP)\pi\pi$ process.

\subsection{Scalar four-point integrals}
\begin{figure}[t]
\begin{center}
\includegraphics[width=0.9\linewidth]{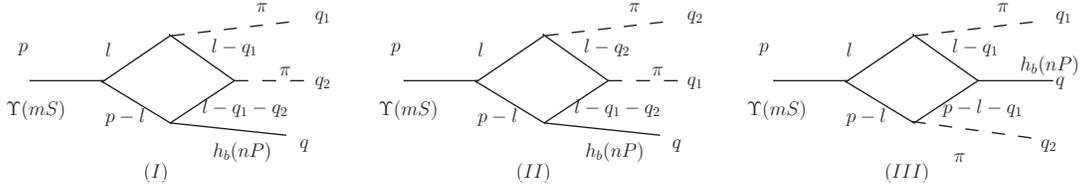}
\caption{Kinematics used in the calculation of the four-point integrals. }
\label{fig:BoxLabel}
\end{center}
\end{figure}

For the first topology as shown in Fig.~\ref{fig:BoxLabel}, the scalar integral evaluated for the initial bottomonium at rest ($p=(M,\vec 0)$) reads
{\allowdisplaybreaks
\begin{align}
 J_1^{(0)}&\equiv i\int \!\frac{d^4l}{(2\pi)^4}\frac{1}{[l^2-m_1^2+i\epsilon][(p-l)^2-m_2^2+i\epsilon][(l-q_1-q_2)^2-m_3^2+i\epsilon][(l-q_1)^2-m_4^2+i\epsilon]}\nonumber\\
&\simeq\frac{-i}{16m_1m_2m_3m_4}\int \! \frac{d^4l}{(2\pi)^4}
\frac{1}{\left[l^0-\frac{\vec l^2}{2m_1}-m_1+i\epsilon\right]\left[l^0-M+\frac{\vec l^2}{2m_2}+m_2-i\epsilon\right]} \nonumber\\
&\times \frac{1}{\left[l^0-q_1^0-q_2^0-\frac{(\vec l+\vec q)^2}{2m_3}-m_3+i\epsilon\right]\left[l^0-q_1^0-\frac{(\vec l-\vec q_1)^2}{2m_4}-m_4+i\epsilon\right]} .
\end{align}}%
Performing the contour integration, we find
\begin{equation}
-\frac{\mu_{12}\mu_{23}\mu_{24}}{2m_1m_2m_3m_4}\int \! \frac{d^3l}{(2\pi)^3}\frac{1}{
[\vec l^2+c_{12}-i\epsilon][\vec l^2+2\frac{\mu_{23}}{m_3}\vec l\cdot \vec q+c_{23}-i\epsilon][\vec l^2-2\frac{\mu_{24}}{m_4}\vec l\cdot \q_1+c_{24}-i\epsilon]} , \label{eq:topo1}
\end{equation}
where we defined
\begin{align}
 c_{12} &\equiv 2\mu_{12}\left(m_1+m_2-M\right), \quad c_{23}\equiv 2\mu_{23}\left(m_2+m_3-M+q_1^0+q_2^0+\frac{\q^2}{2m_3}\right), \nonumber\\
 c_{24} &\equiv 2\mu_{24}\left(m_2+m_4-M+q^0_1+\frac{\q_1^2}{2m_4}\right), \quad \mu_{ij}=\frac{m_im_j}{m_i+m_j}.
\end{align}
The second topology in Fig.~\ref{fig:BoxLabel} is just the crossed diagram of the first topology with $q_1 \leftrightarrow q_2$, so the
scalar integral reads
\begin{equation}
J_2^{(0)}=-\frac{\mu_{12}\mu_{23}\mu_{24}}{2m_1m_2m_3m_4}\int \! \frac{d^3l}{(2\pi)^3}\frac{1}{
[\vec l^2+c_{12}-i\epsilon][\vec l^2+2\frac{\mu_{23}}{m_3}\vec l\cdot \vec q+c_{23}-i\epsilon][\vec l^2-2\frac{\mu_{24}}{m_4}\vec l\cdot \q_2+c_{24}^\prime-i\epsilon]} , \label{eq:topo2}
\end{equation}
where
\begin{align}
 c_{24}^\prime &\equiv 2\mu_{24}\left(m_2+m_4-M+q^0_2+\frac{\q_2^2}{2m_4}\right)\,.
\end{align}


For the third topology we have
\begin{align}
 J_3^{(0)}&\equiv i\int \!\frac{d^4l}{(2\pi)^4}\frac{1}{[l^2-m_1^2+i\epsilon][(p-l)^2-m_2^2+i\epsilon][(p-q_2-l)^2-m_3^2+i\epsilon][(l-q_1)^2-m_4^2+i\epsilon]}\nonumber\\
&\simeq\frac{-i}{16m_1m_2m_3m_4}\int \! \frac{d^4l}{(2\pi)^4}
\frac{1}{\left[l^0-\frac{\vec l^2}{2m_1}-m_1+i\epsilon\right]\left[l^0-M+\frac{\vec l^2}{2m_2}+m_2-i\epsilon\right]}\nonumber\\
&\times \frac{1}{\left[l^0+q_2^0-M+\frac{(\vec l+\vec q_2)^2}{2m_3}+m_3-i\epsilon\right]\left[l^0-q_1^0-\frac{(\vec l-\vec q_1)^2}{2m_4}-m_4+i\epsilon\right]}.
\end{align}
Performing the contour integration, we find
\begin{align}
-\frac{\mu_{12}\mu_{34}}{2m_1m_2m_3m_4}\int \! \frac{d^3l}{(2\pi)^3}\frac{1}{
[\vec l^2+d_{12}-i\epsilon][\vec l^2-2\frac{\mu_{34}}{m_4}\vec l\cdot \vec q_1-2\frac{\mu_{34}}{m_3}\vec l\cdot \vec q_2+d_{34}-i\epsilon]}\nonumber\\
\times\left[\frac{\mu_{24}}{[\vec l^2-2\frac{\mu_{24}}{m_4}\vec l\cdot \q_1+d_{24}-i\epsilon]}+\frac{\mu_{13}}{[\vec l^2+2\frac{\mu_{13}}{m_3}\vec l\cdot \q_2+d_{13}-i\epsilon]} \right], \label{eq:topo2}
\end{align}
where we defined
\begin{align}
 d_{12}&\equiv 2\mu_{12}\left(m_1+m_2-M\right), & d_{34}&\equiv 2\mu_{34}\left(m_3+m_4-q^0+\frac{\q_1^2}{2m_4}+\frac{\q_2^2}{2m_3}\right), \nn\\
 d_{24}&\equiv 2\mu_{24}\left(m_2+m_4-M+q_1^0+\frac{\q_1^2}{2m_4}\right), & d_{13}&\equiv 2\mu_{13}\left(m_1+m_3-M+q_2^0+\frac{\q_2^2}{2m_3}\right).
\end{align}

In all the three cases the remaining three-dimensional momentum integration will be carried out numerically.

\subsection{Tensor reduction}
%

Since the $\Upsilon B^{(*)}\bar{B}^{(*)}$ vertex scales with the momentum of the bottom meson pair, for topology I we have to deal with
\begin{equation}
\frac{-\mu_{12}\mu_{23}\mu_{24}}{2m_1m_2m_3m_4}\int \!
\frac{d^3l}{(2\pi)^3}\frac{f(l)}{ [\vec l^2+c_{12}-i\epsilon][\vec
l^2+2\frac{\mu_{23}}{m_3}\vec l\cdot \vec q+c_{23}-i\epsilon][\vec
l^2-2\frac{\mu_{24}}{m_4}\vec l\cdot \vec q_1+c_{24}-i\epsilon]},
\end{equation}
where $f(l)=\{1,\, l^i\}$ for the fundamental scalar and
vector integrals, respectively. A convenient parametrization of the tensor reduction
reads
\begin{align} \label{eq.TensorReductionOfVectorIntegrals}
J_1^{(1)i}&=\frac{-\mu_{12}\mu_{23}\mu_{24}}{2m_1m_2m_3m_4}\int \! \frac{d^3l}{(2\pi)^3}\frac{l^i}{[\vec l^2+c_1-i\epsilon][\vec l^2-2\frac{\mu_{23}}{m_3}\vec l\cdot \vec q+c_2-i\epsilon][\vec l^2-2\frac{\mu_{24}}{m_4}\vec l\cdot \q_1+c_3-i\epsilon]} \nn\\
&\equiv q^iJ_1^{(1)}+q_{1\perp}^iJ_1^{(2)}\,,
\end{align}
where $\vec q_{1\perp}=\vec q_1-\vec
q(\vec q\cdot \vec q_1)/\vec q^2$. The expressions of the scalar integrals $J^{(r)}_1$ can easily be disentangled and have to be evaluated numerically. The corresponding expressions for topology II and III can be obtained by changing the denominators accordingly.
\subsection{Amplitudes}

We define the scalar integrals $J1(i,r,k)$ based on the $J_1^{(r)}$ in the tensor reduction of vector integral in Eq.~\eqref{eq.TensorReductionOfVectorIntegrals}, where $i=1,2,3$ denotes the three topologies of box diagrams as shown in Fig.~\ref{fig:BoxLabel}, $r=1,2$ refers to
the two components $J_1^{(r)}$, and $k=1,2,...,12$ represents the twelve patterns with different
pseudoscalar or vector content of the intermediate bottom mesons in $[M1,M2,M3,M4]$ as displayed in Sec.~\ref{AmplitudesCalculation}.

We give the amplitude of the box diagrams for the $\Upsilon(mS) \to h_b(nP)\pi\pi$ process, namely the $A_l (l=1,2,...,6)$ in the
Eq.~\eqref{eq.Mloop}.

\bqa
A_1=&&\frac{8 g_1 g_{JHH}g_\pi^2}{F_\pi^2 \q^2}
\bigg\{\q^2 \Big\{\vec p_c\cdot \vec p_d \big[J1(1,1,3)+J1(2,1,3)+J1(3,1,8)\big]+\vec p_c\cdot \vec q \big[J1(1,1,9)+J1(1,1,11)\nn \\&& -J1(2,1,12)+J1(3,1,9)+J1(3,1,11)\big]+\qperp^2 \big[J1(1,2,9)+J1(1,2,11)-J1(2,2,12)+J1(3,2,9)\nn \\&&
+J1(3,2,11)\big]\Big\}+\vec p_c\cdot \vec q \Big\{\vec p_c\cdot \vec q \big[J1(1,1,9)+J1(1,1,11)-J1(2,1,12)+J1(3,1,9)+J1(3,1,11)\big]\nn \\&&
+\vec p_d\cdot \vec q \big[J1(1,1,12)-J1(2,1,9)-J1(2,1,11)+J1(3,1,10)\big]+\qperp^2 \big[J1(1,2,9)+J1(1,2,11)\nn \\&& -J1(1,2,12)+J1(2,2,9)+J1(2,2,11)-J1(2,2,12)+J1(3,2,9)-J1(3,2,10)+J1(3,2,11)\big]\Big\}\bigg\}\,,
\eqa

\bqa
A_2=&&\frac{8 g_1 g_{JHH}g_\pi^2}{F_\pi^2}
\Big\{\vec p_c\cdot \vec p_d \big[J1(1,2,3)+J1(2,2,3)+J1(3,2,8)\big]+\vec p_c\cdot \vec q \big[J1(1,1,9)+J1(1,1,11)\nn \\&&
-J1(2,1,12)+J1(3,1,9)+J1(3,1,11)\big]+\vec p_d\cdot \vec q \big[J1(1,1,12)-J1(2,1,9)-J1(2,1,11)\nn \\&& +J1(3,1,10)\big]+\qperp^2\big[J1(1,2,9)+J1(1,2,11)-J1(1,2,12)+J1(2,2,9)+J1(2,2,11)\nn \\&&
-J1(2,2,12)+J1(3,2,9)-J1(3,2,10)+J1(3,2,11)\big]\Big\}\,,
\eqa

\bqa
A_3=&&-\frac{8 g_1 g_{JHH}g_\pi^2}{F_\pi^2 \q^2}
\Big\{\q^2 \big[-J1(1,1,9)+J1(1,1,11)-J1(1,1,12)-J1(1,2,2)+J1(1,2,9)+J1(1,2,10)\nn \\&&
+J1(1,2,12)-J1
(2,1,9)+J1(2,1,11)-J1(2,1,12)+J1(2,2,2)-J1(2,2,10)-J1(2,2,11)\nn \\&&
-J1(3,1,9)-J1(3,1,10)+J1(3,1,11)-J1(3,2,2)+J1(3,2,9)+J1(3,2,10)+J1
(3,2,12)\big]\nn \\&&
+\vec p_c\cdot \vec q \big[J1(1,2,9)-J1(1,2,11)+J1(1,2,12)+J1(2,2,9)-J1(2,2,11)+J1(2,2,12)\nn \\&&
+J1(3,2,9)+J1(3,2,10)-J1(3,2,11)\big]\Big\}\,,
\eqa

\bqa
A_4=&&\frac{8 g_1 g_{JHH}g_\pi^2}{F_\pi^2\q^4}
\Big\{\q^4 \big[J1(1,1,2)-J1(1,1,10)-J1(1,1,11)-J1(2,1,2)+J1(2,1,9)+J1(2,1,10)\nn \\&&
+J1(2,1,12)+J1
(3,1,2)-J1(3,1,11)-J1(3,1,12)\big]+\q^2 \vec p_c\cdot \vec q \big[J1(1,1,9)-J1(1,1,11)\nn \\&&
+J1(1,1,12)-J1(1,2,9)+J1(1,2,11)-J1(1,2,12)+J1
(2,1,9)-J1(2,1,11)+J1(2,1,12)\nn \\&&
-J1(2,2,9)+J1(2,2,11)-J1(2,2,12)+J1(3,1,9)+J1(3,1,10)-J1(3,1,11)-J1(3,2,9)\nn \\&&
-J1(3,2,10)+J1(3,2,11)\big)-
(\vec p_c\cdot \vec q)^2 \big[J1(1,2,9)-J1(1,2,11)+J1(1,2,12)+J1(2,2,9)\nn \\&&
-J1(2,2,11)+J1(2,2,12)+J1(3,2,9)+J1(3,2,10)-J1(3,2,11)\big]\Big\}\,,
\eqa

\bqa
A_5=&&\frac{8 g_1 g_{JHH}g_\pi^2}{F_\pi^2\q^2}
\Big\{\q^2 \big[-J1(1,1,6)+J1(1,2,8)+J1(1,2,10)-J1(2,1,6)+J1(2,2,6)-J1(2,2,8)\nn \\&&
-J1(2,2,10)-J1
(3,1,3)+J1(3,1,6)-J1(3,1,8)+J1(3,2,3)+J1(3,2,12)\big]+\vec p_c\cdot \vec q \nn \\&&
\big[J1(1,2,6)+J1(2,2,6)+J1(3,2,3)-J1(3,2,6)+J1(3,2,8)\big]\Big\}\,,
\eqa

\bqa
A_6=&&-\frac{8 g_1 g_{JHH}g_\pi^2}{F_\pi^2\q^4}
\Big\{\q^4 \big[J1(1,1,6)-J1(1,1,8)-J1(1,1,10)+J1(2,1,8)+J1(2,1,10)-J1(3,1,6)\nn \\&&
+J1(3,1,8)-J1
(3,1,12)\big]+\q^2 \vec p_c\cdot \vec q \big[J1(1,1,6)-J1(1,2,6)+J1(2,1,6)-J1(2,2,6)+J1(3,1,3)\nn \\&&
-J1(3,1,6)+J1(3,1,8)-J1(3,2,3)+J1
(3,2,6)-J1(3,2,8)\big])-(\vec p_c\cdot \vec q)^2 \big[J1(1,2,6)+J1(2,2,6)\nn \\&&
+J1(3,2,3)-J1(3,2,6)+J1(3,2,8)\big]\Big\}\,.
\eqa

\end{appendix}

\end{document}